\documentclass[journal,twoside,web]{ieeecolor}
\usepackage{generic}
\usepackage{cite}
\usepackage{amsmath,amssymb,amsfonts}
\usepackage{algorithmic}
\usepackage{graphicx}
\usepackage{algorithm,algorithmic}
\usepackage[hidelinks]{hyperref}
\usepackage{textcomp}
\usepackage{subcaption}
\usepackage{multirow}
\graphicspath{{graphics/}}
\DeclareGraphicsExtensions{.pdf}
\usepackage{booktabs}

\newcommand{\TBP}{TB$^+$}

\def\BibTeX{{\rm B\kern-.05em{\sc i\kern-.025em b}\kern-.08em
    T\kern-.1667em\lower.7ex\hbox{E}\kern-.125emX}}
\begin{document}
\title{Deep Learning for Tuberculosis Screening in a High-burden Setting using Cough Analysis and Speech Foundation Models}
\author{Ning Ma, \IEEEmembership{Member, IEEE}, Bahman Mirheidari, Guy J. Brown, Nsala Sanjase, Minyoi M. Maimbolwa, Solomon Chifwamba, Seke Muzazu, Monde Muyoyeta, and Mary Kagujje
\thanks{Ning Ma, Bahman Mirheidari, and Guy J. Brown are with the School of Computer Science, University of Sheffield, Sheffield S1 4DP, UK (e-mail: \{n.ma, b.mirheidari, g.j.brown\}@sheffield.ac.uk).}
\thanks{Nsala Sanjase, Minyoi M. Maimbolwa, Solomon Chifwamba, Seke Muzazu, Monde Muyoyeta, and Mary Kagujje are with Centre for Infectious Disease Research in Zambia, Lusaka, 10101, Zambia  (e-mail: \{nsala.sanjase, minyoi.maimbolwa, solomon.chifwamba, seke.muzazu, monde.muyoyeta, mary.kagujje\}@cidrz.org).}
}
\maketitle
\begin{abstract}
Artificial intelligence (AI) systems can detect disease-related acoustic patterns in cough sounds, offering a scalable and cost-effective approach to tuberculosis (TB) screening in high-burden, resource-limited settings. Previous studies have been limited by small datasets, under-representation of symptomatic non-TB patients, and recordings collected in controlled environments. In this study, we enrolled 512 participants at two hospitals in Zambia, categorised into three groups: bacteriologically confirmed TB (\TBP), symptomatic patients with other respiratory diseases (OR), and healthy controls (HC). Usable cough recordings with demographic and clinical data were obtained from 500 participants. Deep learning classifiers based on pre-trained speech foundation models were fine-tuned on cough recordings to predict diagnostic categories. The best-performing model, trained on 3-second audio clips, achieved an AUROC of 85.2\% for distinguishing TB coughs from all other participants (\TBP/ Rest) and 80.1\% for \TBP~versus symptomatic OR participants (\TBP/ OR). Incorporating demographic and clinical features improved performance to 92.1\% for \TBP/ Rest and 84.2\% for \TBP/ OR. At a probability threshold of 0.38, the multimodal model reached 90.3\% sensitivity and 73.1\% specificity for \TBP/ Rest, meeting WHO target product profile benchmarks for TB screening. Adversarial testing and stratified analyses shows that the model was robust to confounding factors including background noise, recording time, and device variability. These results demonstrate the feasibility of cough-based AI for TB screening in real-world, low-resource settings.
\end{abstract}


\section{Introduction}

Despite being preventable and curable, tuberculosis (TB) remains a major global health challenge, with 10.6 million new cases and 1.3 million deaths reported in 2023~\cite{WHO2023TBreport}. A significant proportion of TB-related deaths, nearly one quarter of all estimated TB cases, stems from undiagnosed and untreated cases~\cite{WHO2023TBreport}. Systematic screening is essential to facilitate earlier diagnosis, reduce community transmission, and improve treatment outcomes~\cite{WHO2021TB}, but current World Health Organisation (WHO)-approved screening tools are limited. Symptom-based screening offers only moderate sensitivity (71\%) and specificity (64\%)~\cite{burke2021community}, while chest X-ray (CXR) requires costly infrastructure, with limited availability in many low-resource settings. C-reactive protein (CRP) testing shows diagnostic utility primarily among people living with HIV (PLHIV). Molecular WHO-recommended rapid diagnostic (mWRD), although valuable, are constrained by moderate sensitivity (69\%), which is only recommended for adolescents and adults living with HIV and requires a robust sample transportation network~\cite{macpherson2025policies, Fleming2021, kiguli2020audit}.

Advances in artificial intelligence (AI) offer opportunities to improve TB screening and diagnosis by enhancing accuracy, efficiency, and accessibility across various healthcare domains, including medical imaging~\cite{Suzuki2017, Qin2021TB, Moodley2022}, disease detection~\cite{Lang2023, Noor2020, Spada2024, Wang2019}, and sleep-disordered breathing monitoring~\cite{Romero2022}. AI-based CXR interpretation tools were recommended by WHO in 2021 as alternatives to human readers for TB screening~\cite{WHO2021TB}, but their use remains constrained by equipment availability~\cite{Fleming2021, kiguli2020audit}. These challenges highlight the need for alternative AI applications in TB screening~\cite{WHO2021TB}, particularly those that exploit other diagnostic substrates such as audio signals and wearable sensor data, which may be more accessible and scalable in low-resource settings. 

An emerging alternative for TB screening is AI-based analysis of cough sounds. Cough is a hallmark symptom of TB, caused by inflammation of the airways, and is thought to contain disease-specific acoustic features distinguishable from coughs of other causes~\cite{Botha2018TBcough}. Clinicians already use cough characteristics such as tone, pitch, intensity, and duration of coughs, alongside associated symptoms, to guide diagnosis~\cite{amos2017cough, Koo2021}. AI-enabled cough analysis could formalise this approach, reduce variability, and scale to low-resource settings~\cite{Smith2006CoughSound}. Recent studies have explored AI-enabled cough sound analysis as a screening tool for TB~\cite{Botha2018TBcough, Pahar2021TBcough, Yellapu2023, Sharma2024TBcough}. However, prior research in this area has been limited by several weaknesses, including small sample sizes~\cite{Botha2018TBcough, Pahar2021TBcough}, data biases due to inadequate inclusion of symptomatic individuals without TB~\cite{Botha2018TBcough, Pahar2021TBcough, Yellapu2023}, the use of relatively simple machine learning models~\cite{Botha2018TBcough, pathri2022}, and data collected in controlled, quiet environments that do not reflect real-world conditions~\cite{Botha2018TBcough, Sharma2024TBcough}.

In this study, we propose a novel AI-enabled cough sound analysis model specifically designed to overcome these limitations. We hypothesise that, with a sufficiently large and well-balanced dataset, machine learning models could learn to recognise spectral and temporal features in cough sounds that reliably distinguish TB from non-TB cases. Our contributions are threefold: (1) we present one of the largest and most balanced datasets of cough sounds for TB screening collected in noisy, real-world environments; (2) we demonstrate the effectiveness of speech foundation models for this task, achieving performance that meets WHO target product profile benchmarks; and (3) we systematically analyse robustness across devices, subgroups, and confounding factors, highlighting pathways for practical deployment in resource-limited settings. Our objective is to assess model performance in a high-burden setting, thereby advancing the development of practical and scalable tools for TB screening. Such tools could be particularly valuable in resource-limited health systems and in high-throughput contexts such as immigration screening centres.

\section{Methods}

\subsection{Study design and participants}

This cross-sectional study follows the Standards for Reporting of Diagnostic Accuracy Studies (STARD) guidelines~\cite{Bossuyt2003}. Ethical approval was obtained from the University of Zambia Biomedical Research Ethics Committee (approval number 3648-2023). Written informed consent was obtained from all participants prior to enrolment.

Adults aged 18 years or older were recruited into three groups: (1) individuals with bacteriologically confirmed TB (TB$^{+}$), defined by a positive Xpert MTB/RIF result; (2) symptomatic patients with respiratory disease but no TB (other respiratory diseases, OR); and (3) asymptomatic healthy controls (HC). Exclusion criteria for the TB$^{+}$ group included prior TB history, ongoing anti-TB treatment for more than three days, or a trace call result on Xpert MTB/RIF testing. TB was excluded in the OR group using sputum Xpert testing and chest X-rays. In the HC group, sputum Xpert MTB/RIF testing was performed.

We aimed to recruit 550 participants (250 TB$^{+}$, 150 OR, and 150 HC). This sample size was chosen to exceed those used in previous studies on AI-driven acoustic analysis for TB screening~\cite{Botha2018TBcough, Pahar2021TBcough, pathri2022, Yellapu2023}, as machine learning models typically benefit from larger and more diverse training datasets~\cite{Ajiboye2015}. In total 512 participants were enrolled between April 2023 and August 2024, from two Level-1 hospitals in Lusaka, Zambia (Kanyama and Chawama), which serve communities with a high burden of both TB and HIV. TB$^{+}$ participants were enrolled consecutively from TB clinics at the study sites. OR participants were recruited from symptomatic patients presenting to outpatient departments, and the HC group was composed of asymptomatic individuals, including caregivers and healthcare workers. To minimise confounding, the OR and HC groups were frequency-matched to the TB$^{+}$ group by age and gender. Gender distribution reflected Zambia's TB notification trends, with men comprising ~64\% of TB cases in 2023 \cite{WHO2023TBreport}. In addition, participant recruitment was balanced across sites and time points to reduce bias from temporal or site-specific factors.

\subsection{Procedures}

All participants underwent a brief clinical evaluation, including medical history and physical examination. Data were recorded on paper forms and later digitised in a secure electronic database.

Cough recordings were captured in sound-attenuated outdoor Keter sheds using identical setups across sites. The sheds were foam-lined to reduce acoustic reflections. The Kanyama shed was located approximately 100\,m from road traffic and situated near the chest clinic. The Chawama shed was similarly distanced from road traffic near the rear hospital gate and adjacent to a church. A high-fidelity stereo microphone pair (R{\O}DE M5) was positioned 50~cm in front of the seated participant, at head height. Simultaneous recordings were also captured on two smartphones (Samsung Galaxy series) placed on the table top in front of the participant. As part of the infection control measures, the microphone was covered by a cut out from a disposable gown which was replaced daily. TB$^+$ participants were recorded after non-TB participants. UV light disinfection was applied after each TB$^+$ recording session for a minimum of 15 minutes.

All recordings were conducted under the guidance of trained research personnel. Each participant provided at least three voluntary cough sessions (2–3 coughs/session). A custom web-based application was used to synchronise and manage sound recordings from all devices. Audio files were saved temporarily on a laptop and subsequently uploaded to a secure cloud storage platform.

\subsection{Model development}

All audio recordings, captured via both condenser microphones and smartphones, were downsampled to 16\,kHz mono and trimmed to exclude silence. Cough segments were automatically extracted using an energy-based detector, with 200\,ms of leading and trailing signals retained to reserve acoustic context. Each sample was labelled according to the participant's diagnostic group: TB$^+$, OR, or HC. 

\begin{figure*}[htb]
\centering
\includegraphics[width=.83\textwidth]{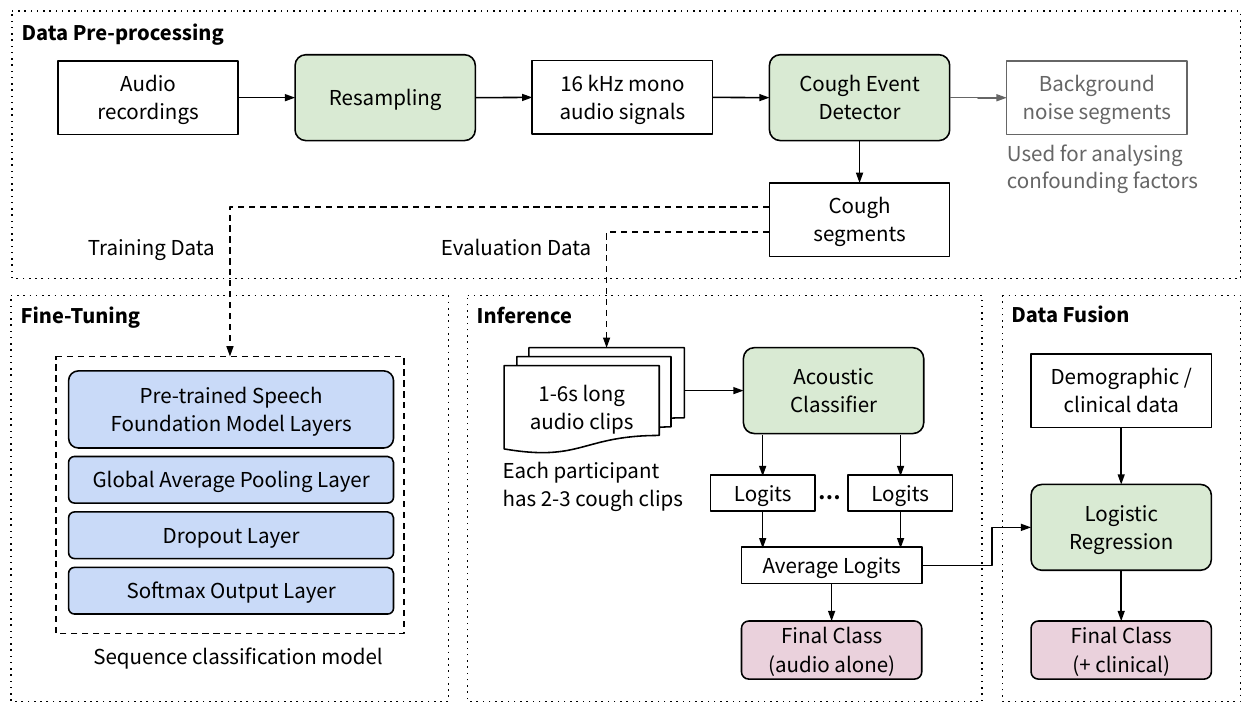}
\caption{\textbf{Pipeline of the automatic cough-based TB screening system using foundation models.}}
\label{fig:pipeline}
\end{figure*}

A multi-stage classification pipeline (Fig.~\ref{fig:pipeline}) was developed using state-of-the-art pre-trained speech foundation models. These models, including Wav2Vec2~\cite{Baevski2020wav2vec}, WavLM \cite{chen2022wavlm}, HuBERT \cite{hsu2021hubert}, Data2Vec \cite{baevski2022data2vec}, and Whisper~\cite{Radford2023robust}, accept raw audio waveform as input and generate contextualised audio embeddings that capture complex acoustic patterns. To adapt each model for three-way classification (\TBP, OR, HC) in this study, we appended a classifier head comprising a global average pooling layer, dropout (rate = 0.5), and a softmax output layer. Fine-tuning was conducted using stratified 10-fold cross-validation with the following parameters: 5 training epochs, batch size = 8, learning rate = 3e-5, warm-up ratio = 0.1, and gradient accumulation over 8 steps. 

Participant-level predictions were derived using a segment-wise soft voting strategy: softmax probabilities from each cough segment of the same participant were averaged, and the class with the highest cumulative probability was assigned as the final prediction. To improve classification performance, we implemented an ensemble stacking approach~\cite{zhou2025ensemble} to combine cough sound information with demographic and clinical data. For each participant, we concatenated the softmax logits (unnormalised scores for each classification category) from the acoustic models with demographic metadata (age, gender, BMI, symptoms) to form a joint feature vector. A logistic regression (LR) model was then trained as a meta-classifier to integrate these features and generate the final prediction.

To maximise data variability and ensure model generalisability, training included both microphone and smartphone recordings. For evaluation, performance was stratified by recording device type: (1) high-fidelity microphones and (2) smartphones. This reflects practical deployment scenarios -- clinical settings using booth microphone recordings versus community settings using mobile devices. 

\subsection{Experimental setup}
As a baseline, we trained a logistic regression model similar to \cite{Pahar2021TBcough} using mel-frequency cepstral coefficient (MFCC) features (40 coefficients plus delta and delta-delta) extracted from cough segments. Hyperparameters were optimised using a grid search across 100 runs, varying regularisation strength (C=1e-6 to 100), solver (lbfgs, saga, liblinear, newton-cg, sag), tolerance (1e-6 to 1e-4), and maximum iterations (1,000 to 10,000).

To assess model robustness, we conducted several supplementary experiments. First, we evaluated model performance across varying cough segment durations (1–6\,s). Second, we performed adversarial testing using alternative acoustic inputs: (a) white noise, (b) non-cough noise segment, and (c) mismatched training/testing data samples (trained on coughs, tested on noise). Noise segments were sourced from non-cough portions of the same recordings to mimic real-world acoustic conditions and assess overfitting to background noise. These analyses were designed to distinguish signal reliance on pathological acoustic biomarkers from confounding artefacts. 

Finally we examined the effect of varying classifier thresholds (range: 0.35 -- 0.55) on model outputs. To compute confidence intervals (CI) for all key performance metrics using audio and clinical features, we performed 10,000 bootstrap resamples of model predictions and calculated the corresponding performance metrics each time.

\subsection{Evaluation metrics}


Model performance was evaluated using area under the receiver operating characteristic curve (AUROC) across three diagnostic tasks: distinguishing \TBP~from all other (OR + HC) participants (\TBP/ Rest), from other respiratory conditions (\TBP/ OR), and from healthy controls (\TBP/ HC). Additional metrics included sensitivity, specificity, positive predictive value (PPV), negative predictive value (NPV), and F1-score. The best-performing model (on high-fidelity recordings) was used for further subgroup analyses, including stratification by HIV status and time of recording. Final inference was also performed on smartphone recordings to simulate mobile deployment performance.


\begin{table*}[thb]
\caption{Demographic and clinical characteristics of participants included in this study, including percentages of data groups, averages with standard deviations, and data ranges.}
    \label{t:demographics}
    \centering
    \begin{tabular}{llllllllllll}
    \toprule
    & \multicolumn{2}{l}{\textbf{TB$^+$}} && \multicolumn{2}{l}{\textbf{OR}}&& \multicolumn{2}{l}{\textbf{HC}}&& \multicolumn{2}{l}{\textbf{Total}}\\ \midrule
    
    \textbf{Total Participants} & 201  &  && 150  &  && 149 &  && 500 \\ \midrule
    \textbf{\quad Males} & 155 & 77\%  && 96 & 64\%&& 90 & 60\%&& 341 & 68\%\\ 
    \textbf{\quad Females} & 46 & 23\% && 54 & 36\% && 59 & 40\%&& 159 & 32\% \\ \midrule
    \textbf{Age} (years) & 34 $\pm$ 10 & 18 -- 73 && 37 $\pm$ 13 & 18 -- 71 && 32 $\pm$ 11 & 18 -- 73 && 34 $\pm$ 11 & 18 -- 73  \\ \midrule
    \textbf{BMI} (kg/m$^2$) & 19 $\pm$ 3 & 14 -- 29 && 22 $\pm$ 4 & 14 -- 40 && 24 $\pm$ 6 & 15 -- 48 && 21 $\pm$ 5 & 14 -- 48  \\ \midrule
    \textbf{Symptom Presence} & 201  & 100\%&& 150 & 100\%&& 0 &0\%&& 351 & 70\%\\
    \midrule
    \textbf{HIV+} & 63  &31\%&& 51 &34\%&& 18 &12\%&& 132 & 26\% \\ 
    \bottomrule
    \end{tabular}
\end{table*}

\begin{table*}[thb]
\caption{Comparison of AUROC with different classifier architectures and audio input durations. Best results are in bold.}
\begin{centering}
\begin{tabular}{l l l l l l l l l}
\toprule 
\textbf{Classifier} & \textbf{Model} & \textbf{Task} & \multicolumn{6}{l}{\textbf{Audio Input Duration}} \\  
\cmidrule{4-9}
\textbf{Architecture} & \textbf{Size} &  & 1 sec & 2 sec & \textbf{3 sec} & 4 sec & 5 sec & 6 sec \\
 \midrule 
LR & 120 & \TBP / Rest & 76.4\%&79.6\%&78.8\%&75.3\%&75.7\%&70.2\% \\
&  & \TBP / OR & 71.5\%&75.2\%&74.3\%&70.9\%&72.6\%&65.2\% \\
&  & \TBP / HC & 81.3\%&84.0\%&83.3\%&79.7\%&78.9\%&75.2\%\\
\midrule
Whisper & 74M & \TBP / Rest & 81.9\%&82.8\%&83.8\%&82.6\%&83.5\%&83.7\% \\
&  & \TBP / OR & 76.4\%&76.7\%&78.3\%&76.8\%&77.3\%&77.8\% \\
&  & \TBP / HC & 87.5\%&89.0\%&89.4\%&88.4\%&89.8\%&89.7\%\\
 \midrule
\textbf{Wav2Vec2} & 95M & \TBP / Rest & 81.8\%&83.9\%&\textbf{85.2\%}&84.8\%&83.0\%&83.3\% \\
&  & \TBP / OR & 76.4\%&78.7\%&\textbf{80.1\%}&79.5\%&76.9\%&78.0\%\\
&  & \TBP / HC & 87.3\%&89.1\%&90.4\%&90.1\%&89.1\%&88.7\%\\
 \midrule
WavLM & 95M & \TBP / Rest &76.4\%&83.5\%&84.6\%&82.5\%&84.2\%&82.7\% \\
&  & \TBP / OR & 70.7\%&77.2\%&78.8\%&75.9\%&78.2\%&75.8\%\\
&  & \TBP / HC & 82.1\%&89.7\%&90.5\%&89.2\%&90.3\%&89.7\%\\  
 \midrule
HuBERT & 95M & \TBP / Rest &78.9\%&84.1\%&84.8\%&83.6\%&84.1\%&83.5\%\\
&  & \TBP / OR & 72.2\%&78.2\%&78.8\%&77.8\%&78.1\%&77.3\%\\
&  & \TBP / HC & 85.6\%&90.1\%&\textbf{90.9\%$^*$}&89.4\%&90.1\%&89.8\% \\  
 \midrule
Data2Vec & 95M & \TBP / Rest &81.6\%&84.1\%&85.1\%&84.2\%&83.4\%&84.5\% \\
&  & \TBP / OR & 75.6\%&78.6\%&79.9\%&78.5\%&78.4\%&78.9\%\\
&  & \TBP / HC & 87.6\%&89.7\%&90.3\%&90.0\%&88.5\%&90.2\% \\ 
 \bottomrule
\end{tabular}
\par\end{centering}
\label{tab:Scaling-Param}
\end{table*}

\section{Results}


Table~\ref{t:demographics} summarises the demographic and clinical characteristics of participants included in each group. Of the 512 participants enrolled, 12 were excluded due to silent or missing audio from at least one recording source. The final analysis included 500 participants: 201 in the TB$^+$ group, 150 in the OR group, and 149 in the HC group. Participants in the TB$^+$ group were predominantly male (77\%), higher than in the OR (64\%) and HC (60\%) groups. This gender imbalance reflects the known higher prevalence of TB among men and may influence model generalisability. The mean age was comparable across groups, but BMI was lowest in the TB$^+$ group (19$\pm$3\,kg/m$^2$).
For symptom presence, as expected all participants in the TB$^+$ and OR groups reported at least one respiratory symptom, whereas all HC participants were asymptomatic. Out of the 500 participants, 132 had HIV co-infection. HIV was more prevalent in both symptomatic groups, with 31\% in TB$^+$ (63 participants) and 34\% in OR (51 participants), compared to only 12\% in the HC group (18 participants). As HIV can alter immune and respiratory responses, this disparity may affect cough acoustics and classifier performance.

\subsection{Different classifier architectures and audio durations}

Table~\ref{tab:Scaling-Param} presents AUROC performance across different classifier architectures and audio input durations for different classification tasks. All foundation model-based classifiers outperformed the LR baseline. The best overall performance was achieved by Wav2Vec2 using 3-second audio clips (AUROC = 85.2\% for \TBP/ Rest, 80.1\% for \TBP/ OR, and 90.4\% for \TBP/ HC). Other foundation models showed comparable performance, with no statistically significant differences observed between them (paired t-test; $p > 0.5$). For all models, 3-second inputs consistently resulted in the highest AUROC values. Performance with 3-second inputs was significantly better than with 1-second inputs across all foundation models, indicating the importance of temporal information. Longer durations (4–6 seconds) did not provide further gains and led to reduced performance for some models.

\begin{table*}[thb]
\caption{Comparison of performance metrics for different classification tasks using various feature sets. The classifier architecture used here is Wav2Vec2 and the audio input duration is 3 sec.}
\centering
\begin{tabular}{llllllll}
\toprule
\textbf{Task} & \textbf{Features} & \textbf{AUROC} & \textbf{Sensitivity} &\textbf{Specificity} & \textbf{PPV} & \textbf{NPV} & \textbf{F1-score} \\
\midrule
\TBP / Rest 
& Audio Alone       & 85.2\%&82.3\%&73.2\%&82.0\%&73.5\%&82.1\%\\
\cmidrule{2-8}
& Audio + Gender   & 85.7\%&83.9\%&74.1\%&82.8\%&75.6\%&83.3\%\\
& Audio + Age      & 85.9\%&83.9\%&73.6\%&82.5\%&75.5\%&83.2\%\\
& Audio + BMI      &  
88.7\%&84.9\%&78.1\%&85.2\%&77.7\%&85.0\%\\
& Audio + Symptom  & 90.1\%&83.2\%&80.6\%&86.4\%&76.4\%&84.8\% \\
\cmidrule{2-8}
& Audio + All Info &  
92.1\%&85.6\%&83.1\%&88.2\%&79.5\%&86.9\% \\
\midrule
\TBP / OR 
& Audio Alone       & 80.1\%&72.0\%&73.2\%&66.7\%&77.8\%&69.2\% \\
\cmidrule{2-8}
& Audio + Gender   & 80.4\%&76.0\%&74.2\%&68.7\%&80.6\%&72.1\%\\
& Audio + Age      & 81.6\%&76.7\%&73.7\%&68.5\%&80.9\%&72.3\% \\
& Audio + BMI      & 83.6\%&74.0\%&78.2\%&71.7\%&80.1\%&72.8\% \\
& Audio + Symptom  & 80.3\%&66.7\%&80.6\%&72.0\%&76.4\%&69.2\% \\
\cmidrule{2-8}
& Audio + All Info &84.2\%&71.3\%&83.1\%&75.9\%&79.5\%&73.5\%\\

\midrule
\TBP / HC 
& Audio Alone       &  90.4\%&92.6\%&73.1\%&71.8\%&93.0\%&80.9\% \\
\cmidrule{2-8}
& Audio + Gender   & 91.0\%&91.9\%&74.2\%&72.5\%&92.6\%&81.0\%\\
& Audio + Age      & 90.2\%&91.3\%&73.7\%&72.0\%&91.9\%&80.4\% \\
& Audio + BMI      & 93.8\%&96.0\%&78.2\%&76.5\%&96.3\%&85.1\% \\ 
& Audio + Symptom  & 99.9\%&100.0\%&80.7\%&79.3\%&100.0\%&88.4\% \\
\cmidrule{2-8}
& Audio + All Info & 100.0\%&100.0\%&83.1\%&81.4\%&100.0\%&89.7\%\\ 
\midrule
\TBP/ Rest (HIV+) & Audio Alone     & 81.5\%	& 75.4\%& 69.7\% & 	73.2\% & 	72.1\%	& 74.1\% \\ 
\cmidrule{2-8}
 & Audio + All Info     & 91.8\%&81.9\%&90.9\%&94.7\%&71.5\%&87.8\% \\  

\bottomrule
\end{tabular}
\label{tab:w2v_sup_combined}
\end{table*}

\subsection{Incorporating demographic and clinical features}

\begin{figure*}[thb]
\centering
\begin{subfigure}[t]{0.48\textwidth}
    \centering
    \includegraphics[width=\linewidth]{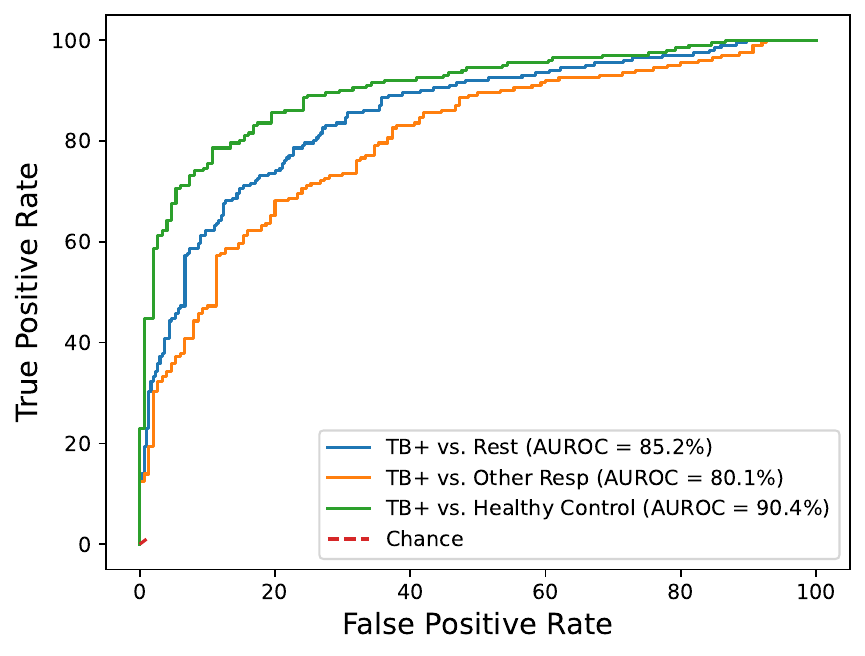}
    \label{fig:wav2vec2-3sec-auc}
\end{subfigure}
\hfill
\begin{subfigure}[t]{0.48\textwidth}
    \centering
    \includegraphics[width=\linewidth]{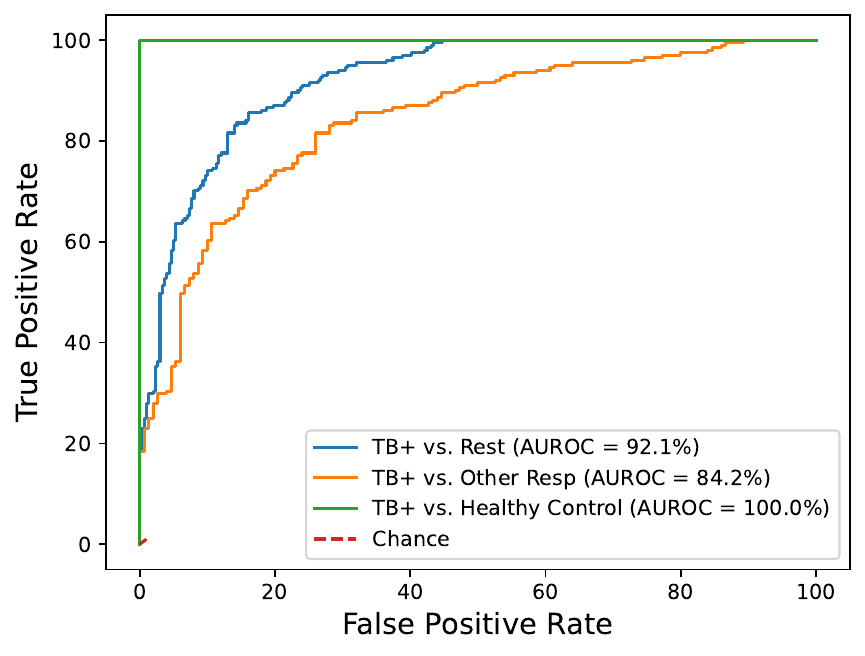}
    \label{fig:d2vec-3sec-auc-info}
\end{subfigure}
\caption{\textbf{Left:} ROC for the Wav2Vec2-based classifier (3 seconds of audio) showing AUC for \TBP~vs. Rest, OR, and HC. \textbf{Right:} ROC for the Wav2Vec2 classifier when all demographic and clinical features were added.}
\label{fig:combined-roc}
\end{figure*}

Using the Wav2Vec2-based classifier and 3-second audio input, we evaluated classification performance when demographic or clinical features (gender, age, BMI, or symptom presence) are available (Table~\ref{tab:w2v_sup_combined}). The ROC curves in different tasks are shown in Fig.~\ref{fig:combined-roc}. With audio alone, classification performance was consistently better for the \TBP/ HC task than for \TBP/ OR. Specificity was similar across both tasks (73.1\% vs. 73.2\%), but sensitivity (92.6\% vs. 72.0\%), PPV (71.8\% vs. 66.7\%), NPV (93.1\% vs. 77.8\%) and F1-score (80.9\% vs. 69.1\%) were all notably higher for \TBP/ HC. These differences suggest greater acoustic overlap between TB and other symptomatic respiratory conditions than between TB and healthy controls. When the OR and HC groups are combined in the \TBP/ Rest task, sensitivity is 82.3\%, specificity is 73.2\%, PPV is 82.0\%, NPV is 73.5\% and F1-score is 82.1\%. Adding individual demographic or clinical features to audio inputs resulted in modest improvements across all tasks. The most substantial gains were observed when all features were combined. For \TBP/ Rest, the inclusion of all supplemental information increased AUROC from 85.2\% to 92.1\%, sensitivity from 82.3\% to 85.6\%, and specificity from 73.2\% to 83.1\%. Similarly, F1-score improved from 82.1\% to 86.9\%. In the more challenging \TBP/ OR task, where both groups were symptomatic, AUROC increased from 80.1\% to 84.2\%, and specificity from 73.2\% to 83.1\%, although sensitivity declined slightly from 72.0\% to 71.3\%. Overall, combining audio with demographic and clinical data led to consistent performance improvements across most metrics, with the strongest effects seen in the \TBP/ Rest task.

\subsection{HIV co-infection}

Results for a subgroup of participants with HIV co-infection (n = 132) are also reported. Among these, 63 were \TBP and 69 were classified as Rest. In this subgroup, classification using audio alone yielded an AUROC of 81.5\%, with sensitivity of 75.4\% and specificity of 69.7\%. When all additional features were included, performance improved markedly, with AUROC reaching 91.8\%, sensitivity 81.9\%, and specificity 90.9\%. 

\subsection{Different classification thresholds}

Table~\ref{tab:final_results_with_CI} shows model performance with different classification thresholds ranging from 0.36 to 0.50. For the \TBP/ Rest task, a threshold of 0.38 offered a strong balance of sensitivity (90.3\%) and specificity (73.1\%), meeting the WHO Target Product Profile for TB screen tools, which recommended a minimum of \textgreater90\% sensitivity and \textgreater70\% specificity for screening tools in 2021~\cite{WHO2021target} and a minimum of \textgreater90\% sensitivity and \textgreater60\% specificity in 2025~\cite{WHO2025target}. Higher thresholds improved specificity (up to 83.1\% at 0.50) at the cost of reduced sensitivity (85.6\% at 0.50). For \TBP/ OR, performance was generally lower than \TBP/ Rest. At the 0.38 threshold, sensitivity was 80.6\% and specificity was 73.1\%. As the threshold increased, specificity improved (up to 83.1\%) while sensitivity dropped (71.3\% at 0.50). In the \TBP/ HC task, the model achieved perfect sensitivity (100\%) across all thresholds tested, with specificity increasing from 70.2\% to 83.1\% between thresholds 0.36 and 0.50.

\begin{table*}[thb]
\caption{Comparison of results for the proposed system (audio + all demographic information) with different threshold scores. Bracketed figures show 95\% confidence intervals based on 10,000 repeats. The results meeting the WHO target profile for TB screening tests are in bold.}
\centering
\begin{tabular}{lllllll}
\toprule
\textbf{Task} & \textbf{Threshold} & \textbf{Sensitivity} & \textbf{Specificity} & \textbf{PPV} & \textbf{NPV} & \textbf{F1-score} \\
\midrule  
\TBP / Rest & \textbf{0.36} & \textbf{91.3\%} & \textbf{70.1\%} & 82\%& 84.4\%& 86.4\%\\ &  & (87.9 to 94.4)& (63.7 to 76.4)& (77.8 to 86)& (78.6 to 89.7)& (83.4 to 89.1)\\ 
&\textbf{0.38}&\textbf{90.3\%}& \textbf{73.1\%}& 83.3\%& 83.6\%& 86.7\%\\ &  & (86.8 to 93.5)& (66.8 to 79.1)& (79.3 to 87.3)& (77.8 to 88.8)& (83.7 to 89.5)\\ 
&0.40&89.3\%& 74.7\%& 84\%& 82.4\%& 86.5\%\\ &  & (85.6 to 92.7)& (68.6 to 80.5)& (79.9 to 88)& (76.6 to 87.8)& (83.5 to 89.3)\\ 
&0.45&87\%& 78.6\%& 85.8\%& 80.2\%& 86.4\%\\ &  & (83.1 to 90.7)& (72.7 to 84.1)& (81.6 to 89.7)& (74.6 to 85.7)& (83.3 to 89.2)\\ 
&0.50&85.6\%& 83.1\%& 88.2\%& 79.5\%& 86.9\%\\ &  & (81.5 to 89.4)& (77.7 to 88.1)& (84.5 to 91.9)& (73.9 to 84.8)& (83.9 to 89.7)\\ 

\midrule
\TBP / OR &0.36&82.7\%& 70.1\%& 67.4\%& 84.4\%& 74.2\%\\ &  & (76.4 to 88.6)& (63.6 to 76.3)& (60.7 to 73.9)& (78.8 to 89.8)& (68.8 to 79.2)\\ 
& 0.38& 80.6\%& 73.1\% & 69.1\%& 83.5\%& 74.4\%\\ &  & (74 to 86.8)& (67 to 79.1)& (62.3 to 75.8)& (77.8 to 88.8)& (68.8 to 79.5)\\ 
&0.40&78.6\%& 74.6\%& 69.8\%& 82.4\%& 73.9\%\\ &  & (71.7 to 85)& (68.6 to 80.4)& (62.8 to 76.7)& (76.7 to 87.7)& (68.3 to 79.1)\\ 
&0.45&74\%& 78.6\%& 72.1\%& 80.2\%& 73\%\\ &  & (66.9 to 80.9)& (72.9 to 84.2)& (64.8 to 78.9)& (74.6 to 85.6)& (67.1 to 78.4)\\ 
&0.50&71.3\%& 83.1\%& 75.9\%& 79.5\%& 73.5\%\\ &  & (63.9 to 78.4)& (77.8 to 88.2)& (68.6 to 83)& (74 to 84.8)& (67.6 to 79)\\ 

\midrule
\TBP / HC &0.36&100\%& 70.2\%& 71.3\%& 100\%& 83.2\%\\ &  & (100 to 100)& (63.8 to 76.4)& (65.2 to 77.4)& (100 to 100)& (78.9 to 87.2)\\ 
& 0.38& 100\%& 73.2\%& 73.4\%& 100\%& 84.6\%\\ &  & (100 to 100)& (67 to 79.2)& (67.2 to 79.4)& (100 to 100)& (80.4 to 88.5)\\ 
&0.40&100\%& 74.6\%& 74.5\%& 100\%& 85.3\%\\ &  & (100 to 100)& (68.4 to 80.5)& (68.4 to 80.4)& (100 to 100)& (81.2 to 89.2)\\ 
&0.45&100\%& 78.6\%& 77.6\%& 100\%& 87.4\%\\ &  & (100 to 100)& (72.6 to 84.1)& (71.5 to 83.5)& (100 to 100)& (83.4 to 91)\\ 
&0.50&100\%& 83.1\%& 81.4\%& 100\%& 89.7\%\\ &  & (100 to 100)& (77.8 to 88.1)& (75.6 to 86.8)& (100 to 100)& (86.1 to 93)\\ 

\bottomrule
\end{tabular}
\label{tab:final_results_with_CI}
\end{table*}

\subsection{Potential confounders} 

\begin{table}[thb]
\centering
\caption{Effect of different noise training and testing conditions using the proposed audio classifier.}
\begin{tabular}{llll}
\toprule 
\textbf{Task} & \textbf{Train Data} & \textbf{Test Data} & \textbf{AUROC} \\
\midrule
\TBP/ Rest & White Noise & White Noise & 56.2\% \\
& Background & Background & 69.9\% \\
& Cough Sound & Background & 58.6\% \\
& Cough Sound & Cough Sound & \textbf{85.2\%} \\
\bottomrule  
\end{tabular}
\label{tab:noise-effect}
\end{table}

When both trained and tested using white noise as audio input, the model performed at around the chance level (AUROC 56.2\%), indicating no meaningful discriminative capacity in the absence of cough sound (Table~\ref{tab:noise-effect}). In contrast, training and testing on non-cough acoustic background led to a significant increase in performance (AUROC = 69.9\%), which suggests the presence of some background noise patterns associated with different participant groups. However, this performance remained well below that observed when trained and tested on cough sounds (AUROC = 85.2\%), confirming that cough acoustics carry the primary discriminatory information. Notably, when the model was trained on cough sounds and tested on background noise, AUROC values again approached chance (58.6\%), further underscoring the specificity of the learned representations to cough content rather than background artefacts.

\begin{figure*}[htb]
\centering
\includegraphics[width=.9\linewidth]{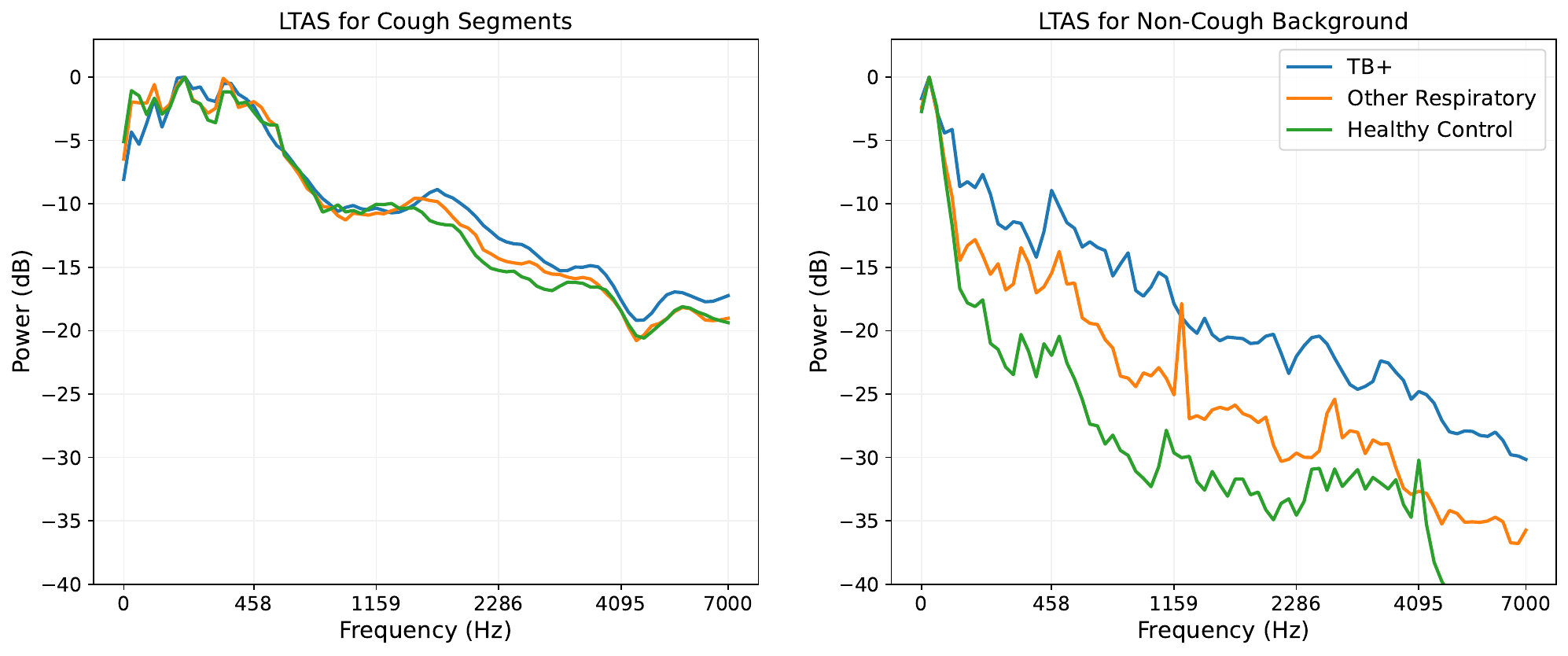}
\caption{Long-term average spectrum (LTAS) of cough audio segments and non-cough background audio.}
\label{fig:ltas}
\end{figure*}

The spectral energy distribution of cough segments and background noise across groups can be visualised in the long-term average spectrum (LTAS) (Fig.~\ref{fig:ltas}). Cough LTAS profiles (left panel) were broadly similar across \TBP, OR, and HC groups, with only minor differences: \TBP~exhibiting slightly lower energy around 200\,Hz and higher energy above 1.5\,kHz. This indicates that group-level differences in cough acoustics are not prominently expressed in the long-term spectral domain, suggesting that temporal features may offer greater diagnostic value. In contrast, background noise LTAS (right panel) showed clearer group differences, with \TBP~recordings exhibiting a flatter, higher-energy spectrum across frequencies, likely due to contextual (e.g., recording environment or timing) rather than physiological factors. However, adversarial testing (Table~\ref{tab:noise-effect}) showed that any such confounding is minimal, as near-chance performance was reported when models were trained on cough sounds and tested on background noise.

\subsection{Mobile recordings}
There was a slight decline in the performance metrics when using mobile phone recordings (Table~\ref{tab:w2v_combined_mobile}). For the \TBP/ Rest classification task, the classifier achieved an AUROC of 83.5\%, which improved to 91.2\% with the inclusion of supplementary information. Similarly, specificity increased from 72.6\% to 80.6\%, and specificity improved from 79.9\% to 84.3\%. The classifier achieved an AUROC of 78.5\% for the \TBP/ OR task using mobile phone recordings, which increased to 82.5\% when supplementary information was included. Sensitivity decreased slightly from 70.7\% to 68.7\% with the additional information, while specificity increased from  72.6\% to 80.6\%.

\begin{table*}[thb]
\caption{Performance of classifiers (with and without additional demographic information) using mobile phone recordings.}
\centering
\begin{tabular}{llllllll}
\toprule
\textbf{Task} & \textbf{Features} & \textbf{AUROC} & \textbf{Sensitivity} & \textbf{Specificity} & \textbf{PPV} & \textbf{NPV} & \textbf{F1-score} \\
\midrule
\TBP / Rest
& Audio Alone               & 83.5\%	& 79.9\%& 72.6\% & 	81.3\% & 	70.9\%	& 80.6\% \\
& Audio + All Info     & \textbf{91.2\%} & \textbf{84.3\%} & \textbf{80.6\%} & \textbf{86.6\%} & \textbf{77.5\%} & \textbf{85.4\%} \\  

\midrule
\TBP / OR 
& Audio Alone               & 78.5\% & \textbf{70.7\%} & 72.6\% & 65.8\% & 76.8\% & 68.1\% \\
& Audio + All Info     & \textbf{82.5\%} & 68.7\% & \textbf{80.6\%} & \textbf{72.6\%} & \textbf{77.5\%} & \textbf{70.5\%} \\ 
\midrule
\TBP / HC 
& Audio Alone           & 88.6\%
& 89.3\%  &72.6\%  & 70.7\%  &90.1\%  & 78.9\%   \\
& Audio + All Info  & \textbf{100.0\%}   &  \textbf{100.0\% } &\textbf{80.6\%} &\textbf{79.2\%}  &\textbf{100.0\%}  &\textbf{88.4\%}\\    
\bottomrule
\end{tabular}
\label{tab:w2v_combined_mobile}
\end{table*}



\section{Discussion}


%
%
This study demonstrates the potential of AI-enabled cough sound analysis using speech foundation models to identify TB from short cough recordings. Our multimodal model achieved performance that meets the WHO target product profile benchmarks for TB screening tools~\cite{WHO2021target, WHO2025target} in distinguishing TB-positive cases from individuals without TB (\TBP/Rest, AUROC 92.1\%, sensitivity 90.3\%, specificity 73.1\%). Based on a well-balanced cohort of 500 participants (201 \TBP, 150 symptomatic non-TB, 149 healthy controls), these findings highlight the promise of cough-based AI as a scalable, non-invasive, and low-cost screening tool in high-burden settings with limited diagnostic infrastructure. 

Compared with previous studies, our model achieved robust and reproducible performance despite real-world recording conditions.
Direct comparisons are limited due to key methodological differences across studies. Prior research in this area has a number of weaknesses, including smaller or imbalanced datasets~\cite{Botha2018TBcough, Pahar2021TBcough}, limited inclusion of symptomatic non-TB patients~\cite{Botha2018TBcough, Pahar2021TBcough, Yellapu2023}, the use of relatively simple machine learning models~\cite{Botha2018TBcough, pathri2022}, and data collection in controlled, quiet environments~\cite{Botha2018TBcough, Sharma2024TBcough}, leading to potential bias and inflated accuracy. Our findings align with real-world clinical practice and prior studies~\cite{kafentzis2023}, demonstrating that the inclusion of demographic and clinical parameters, such as age, gender, BMI, and symptom presence, can enhance model performance. For the \TBP/ Rest task, model sensitivity improved substantially, primarily driven by the inclusion of asymptomatic healthy controls. As this subgroup lacks symptoms, the distinction between TB and non-TB cases becomes clearer, enabling the model to more accurately detect TB-positive individuals. In contrast, for the \TBP/ OR task, where both groups were symptomatic, clinical features contributed primarily to improved specificity, allowing the model to better distinguish TB from other respiratory illnesses.

%
%
The lower performance of the proposed model in the \TBP/ OR task compared to \TBP/ HC highlights the clinical complexity of differentiating TB from other respiratory conditions. Including symptomatic non-TB participants is crucial for realistic rigorous performance assessment. Prior studies have often overlooked this group, or employed a small imbalanced dataset, which could lead to inflated accuracy estimates~\cite{Botha2018TBcough, Pahar2021TBcough, Yellapu2023}. For example, Sharma et al.~\cite{Sharma2024TBcough} reported highly variable classifier performance (AUROC 61–86\%), largely due to training on imbalanced datasets (103 TB patients and 46 non-TB participants). 

%
%
A key technical finding is the importance of temporal context. Across all models, 3-second audio segments consistently outperformed both shorter and longer inputs, which indicates that a brief but temporally rich window best captures relevant cough features. This suggests that TB-related acoustic signatures extend beyond the cough peak but are diluted in longer windows with more irrelevant noise.
This finding contrasts with many earlier studies that used only 0.5--1 second segments centred on the cough peak~\cite{CodaTB2024}, and suggests that TB-related acoustic signatures extend beyond isolated sound peaks.

%
%
Device variability is an unavoidable challenge in real-world deployment. The collected dataset included both smartphone and desktop microphone recordings. Although performance remained strong across both (AUROC 91.2\% on mobile recordings vs. 92.1\% on desktop recordings for \TBP/ Rest), mobile recordings were slightly less accurate, particularly for \TBP/ OR (82.5\% vs. 84.2\%). Domain adaptation and noise-robust training may help mitigate these device variability effects and support community-based deployment.

%
%
Subgroup analysis of 132 HIV-positive participants showed that cough-based classification remains effective for PLHIV, particularly when supplemented with demographic and clinical data. The altered immune response in HIV may cause changes in cough characteristics~\cite{diaz2003respiratory}, but model performance remained comparable to the overall cohort, and surpassed the performance of traditional TB screening tools used in this population, either by achieving higher sensitivity or by offering a better balance between sensitivity and specificity~\cite{WHO2021TB}. This is encouraging, given the high burden of TB-HIV co-infection and the potential for overlapping or atypical symptom presentation. The model's higher performance in PLHIV suggests that their cough patterns were more distinct compared to those in HIV-negative individuals. This could be due to a more homogeneous cough profile in PLHIV, leading to more uniform and recognisable patterns that AI models can learn more effectively. However, as we did not assess immunological markers (e.g., CD4 count) or the nature of lung involvement in the two groups, definitive conclusions cannot be made. Further, the limited sample size in this subgroup (132 participants in total) warrants caution, and additional data are needed to validate model performance in immunocompromised populations.


%
%
We also systematically evaluate confounders such as background noise and recording time. Background noise spectra differed across groups, and simply training and testing on background noise could achieve performance significantly better than chance (AUROC = 69.9\%). 
While there was no significant difference in the hour of recording across groups ($F($2, $N$–3$)$ = 1.59, $p$ = 0.204), we did observe imbalances in the day-wise distribution. Specifically, 48\% of \TBP~participants were recorded on days when only other \TBP~participants were recorded, and a further 23\% on days shared with OR participants. Only 8\% of \TBP~recordings occurred on days shared exclusively with healthy controls. This raises the possibility that the classifier might inadvertently learn spurious correlations associated with recording conditions, such as background noise specific to certain days or locations. However, the classifier's near-chance performance when tested on background noise alone suggests that any such confounding is minimal. 
These findings suggest that our classifier primarily relied on disease-specific cough features rather than environmental cues, which may not have been the case in previous studies. For example, the Swaasa AI study~\cite{Yellapu2023} involved 567 participants, but the symptomatic non-TB group was recorded in a separate study conducted one year earlier, at a different hospital and using different recording equipment~\cite{rudraraju2020cough}. Such separation likely introduced confounding acoustic cues that could be exploited by the classifier to artificially boost performance. By contrast, our data were collected in noisy, real-world conditions using consistent protocols, providing stronger internal validity. Having standard datasets or consistent protocols is therefore crucial in moving performance forward in this area, so that different techniques can be objectively compared.

%
%
This study has several strengths. It involved a large, well-balanced dataset of symptomatic and asymptomatic participants, matched by age, gender, and time of recording. The data was collected in a real-world setting (outdoor and noisy hospitals) which improves its applicability. Additionally, the analysis explored the added value of clinical and demographic features, assessed model performance in HIV-positive individuals, and systematically evaluated potential confounding factors such as background noise, recording time, and device variability.

There are also limitations. First, the study did not include participants with subclinical TB (TB-positive individuals without symptoms), so the model’s ability to detect TB in this subgroup remains unknown. Future studies should assess its performance in individuals with subclinical TB to better understand its real-world screening potential in asymptomatic TB patients. Second, our participants were from a single geographical region, yet AI models have been shown to exhibit regional variations in performance~\cite{mittermaier2023bias}. These differences could be influenced by factors such as air pollution, climate, and the prevalence of other respiratory diseases, which vary across regions and may affect cough characteristics or recording conditions. Lastly, the impact of the speech foundation model size on classification accuracy was not fully explored. A deeper analysis of how model size affects sensitivity, specificity, and computational efficiency could help optimise AI-enabled cough analysis for real-world implementation, particularly in resource-limited settings.

\section{Conclusion}

Our study highlights the potential of AI-enabled cough sound analysis as a scalable and effective TB screening tool, in both PLHIV and HIV negative individuals. The model demonstrated strong performance, with improved accuracy when incorporating clinical features, and its robustness when using both laptop and mobile phone recordings suggests promise for point-of-care applications. AI-enabled cough analysis could enhance TB detection, especially in high-risk populations and resource-limited settings. 

Future research should assess generalisability across regions, extend to subclinical TB, and benchmark performance against chest X-rays or molecular diagnostics. Exploring model compression and training efficiency will also be important for real-world deployment in resource-constrained settings.



\section*{Data sharing}

The TB cough sound dataset and associated code used in this study will be made available to qualified academic and non-profit researchers upon reasonable request following publication, subject to a data use agreement with the University of Sheffield. Data access requests should be submitted to the corresponding author.

\section*{Acknowledgements}

This work was supported by the UK Higher Education Innovation Fund (X/179090) and the UK Engineering and Physical Sciences Research Council (EPSRC) Impact Acceleration Account (R/185787). The authors would like to thank Ms Regina Banda and Ms Winfrida Mashili for their support with data collection, and Ms Sheba Nalwaba for assistance with data entry. We are also grateful to the management and staff of Kanyama and Chawama First-Level Hospitals for providing a supportive environment for data collection.


\bibliographystyle{IEEEtran}
\bibliography{references}

%








\end{document}